# Anisotropy of ocean surface wave turbulence

Alexander M Balk[1]

*Department of Mathematics, University of Utah, 155 South 1400 East, Salt Lake City, Utah 84112*

(*Electronic mail: balk@math.utah.edu)

The paper continues to study the long-standing problem of quasi 1-D (one dimensional) spectrum of sea surface wave turbulence. The study is based on Hasselmann's kinetic equation, which significantly simplifies for the quasi 1-D turbulence. The goal of the paper is to compare the theoretical predictions with experimental observations. The paper starts with *new* derivation of the quasi 1-D approximation of Hasselmann's equation. The present work does not support the imposition of boundary conditions. Instead, the family of theoretical spectra is compared with P-M (Pierson-Moskowitz) and JONSWAP (Joint North Sea Wave Project) spectra. First, a simple solution (first approximation) of the quasi 1-D equation is obtained and compared with the P-M spectrum. Then self-similar solutions are considered; they are shown to satisfy the second order ordinary differential equation. The solutions of this equation have many features of the JONSWAP spectra.

## I. INTRODUCTION

Observing sea waves (gravity waves on sea surface), one immediately notes that the long waves propagate in a single direction. And this happens[1] even though relatively short waves — directly generated by the wind — have a wider span of directions. One can also note that long waves often propagate faster than the wind, and so, the wind would actually act to damp long waves.

This observation seems to contradict our common sense: Longer waves appear due to interaction of shorter waves. Even if the short waves, generated by the wind, had a single direction, they would produce longer waves in a span of directions. Interaction of that longer waves would produce even longer waves, that propagate in a *wider* span of directions. Thus, after a few steps in this inverse cascade process, the turbulence spectrum would be essentially isotropic.

The same isotropization argument is used to obtain the well known Kolmogorov spectrum[2] for the turbulence of incompressible fluid: Eddies aligned in a single direction would produce eddies in a span of directions. So, inside the inertial range, the spectrum should be essentially isotropic. The same argument is also used to justify Ktaichnan's[3] two isotropic spectra of 2-D turbulence (for the inverse and direct cascades).

The spectrum of surface wave turbulence was studied by the JONSWAP team[4]. They actually considered a sharp spectral *peak* (rather than spectrum concentrated around a single direction). Their comprehensive experimental study confirmed that the surface wave turbulence is described by Hasselmann's wave kinetic equation[5,6] (especially for long waves). In particular, they found that wave-wave interactions appeared to account for the observed down-shift of the peak (towards lower frequencies); however, "the origin of the peak and its persistence remained to be explained".

The peak was investigated by Longuet-Higgins[7] in a simplified model based on the Davey-Stewartson equation. He approximated the peak by a Gaussian function (bell-shaped). The results were extended by Fox[8], who approximated the peak by a sum of several Gaussian functions. They found "no indication that a peak could be built up by the weak nonlinear energy transfer", described by Hassselmann's equation.

Zakharov and Smilga[9] studied wave turbulence when its spectrum concentrated along a selected direction (in the wave vector space) instead of the sharp spectral peak. They used Hasselmann's equation and found that it always leads to "a broadening in angle" of the turbulence spectrum. So, they attempted to explain the apparent spectrum anisotropy by non-conservative effects.

In an effort to find the anisotrpization mechanism due to Hasselmann's equation, Irosnikov[10] replaced Hasselmann's equation by its differential approximation and found some evidence for the emergence of turbulence spectra "narrow in angle". However, the presented spectra had a weak anisotropy, in the form of the *first* angular harmonic.

Hasselmann's equation was known to have exact stationary non-equilibrium solutions[11–13]. These solutions — Kolmogorov-Zakharov (KZ) spectra — are isotropic, but might be unstable with respect to anisotropic perturbations. Indeed, it turned out that KZ spectra can be unstable with respect to some high-order angular harmonics (when perturbation is proportional to $e^{i\ell\theta}$, $\theta$ is polar angle, $\ell$ is the order). For example, in some model situation, a KZ spectrum was unstable only to perturbation in the form of the seventh angular harmonic ($\ell = 7$). However, in the case of long surface gravity waves, the KZ spectrum turned out to be stable (both in absolute sense and in the convective sense)[14,15].

The unsuccessful attempts to explain essentially anisotropic spectrum on the basis of Hasselmann's equation lead to considerations of other effects. Studying influence of wind turbulence on ocean surface, Tsimring[16] suggested that formation of narrow angular spectrum could be caused by nonlinear interactions between water waves and turbulence in the air.

A closer agreement with experiments was achieved by Zakharov and Shrira[17]. They considered interaction of surface wave turbulence with the subsurface drift flow (which is always produced when the wind blows long enough).

Still, questions remained. The long wave turbulence is well described by Hasselman's equation. Why does not this equation describe the essential anisotropy of long wave turbulence spectrum? Building upon the previous work[9], Zaslavskii[18–20], considered "the narrow directional approximation", its self-similar solutions, and found "the angular self-compression effect".

Let us mention some further developments: Zakharov, Pushkarev, and Resio[21,22] revisited the differential approximation (mentioned above); Badulin, Pushkarev, Resio, and Zakharov[23,24] studied self-similar solutions of Hasselmann's equation; Badulin and Zakharov[24,25] considered the evolution of anisotropic swell.

It is also worth mentioning, that a similar situation of essentially anisotropic spectrum takes place in the turbulence of Rossby waves or drift waves in plasma. There the dispersion law is anisotropic, but its anisotropy is rather weak (in the form of the first angular harmonic), and so, does not explain the strong anisotropy of turbulence spectra. The spectral anisotropy manifests itself in coordinate space as zonal flow (or alternating zonal jets). However, the Rossby wave turbulence possesses an extra adiabatic invariant (in addition to the energy and momentum); moreover, if the extra invariant is decreasing (due to some external factors), while the energy is not decreasing, then the spectral anisotropy is stronger[26]. Unfortunately, it appears unlikely that a similar invariant could exist for surface gravity waves.

### A.  Results of the present paper

The reported work follows in the steps of Zkharov, Smilga, and Zaslavsky (Refs[9,18–20]) and develops the following picture. The spectral anisotropy grows only in the part of the spectrum close to the downshifting spectral peak. Say, $n(p,q,t)$ is the wave action spectrum of ocean surface turbulence, $t$ is time, $\mathbf{k} = (p, q)$ is the wave vector, with $p$ being its component along the main propagation direction and $q$ the transverse component. Then for any fixed $p$, the anisotropy is decreasing, i.e. the spectrum is broadening in angle (consistent with finding in[9]). However, if we move together with the peak, the spectrum actually becomes narrower in angle, i.e. anisotropy grows. Calculations show that downshifting is critical for the angular narrowing.

The goal of the present work is to investigate how close the theoretical spectrum (resulting from the narrow angle approximation) to the well established experimental JONSWAP spectrum.

The paper presents a new derivation of the narrow angle approximation, which is different from both ones by Zakharov and Smilga[9] and by Zaslavskii[18–20]. I believe the new derivation is simpler and more transparent.

The paper assumes only the angular narrowness and does not assume *apriori* that the spectrum has a peak (with respect to both frequency and angle), but the calculations lead to spectral peak, which has a typical form of the JONSWAP spectrum[4]: Its forward face is "very steep" (the spectrum sharply decreases from the peak value to zero as the frequency $\omega$ decreases below the peak frequency $\omega_m$). On its backward face (when $\omega > \omega_m$), the spectrum decreases gradually.

The paper does not support the imposition of boundary conditions (suggested in[18–20]). The narrow angle approximation seems to break down away from both boundaries ($\omega \to 0$, $\omega \to \infty$). On its forward face (when $\omega \ll \omega_m$), the spectrum changes with $\omega$ too fast, and this change cannot be neglected compared to the sharp change in the transverse direction. (Let alone, the validity of Hasselmann's kinetic equation can fail when the spectrum is changing too fast.) On the backward face (when $\omega \gg \omega_m$), the turbulence spectrum gradually becomes isotropic, before the boundary is reached.

So, we work with the entire family of theoretical spectra (resulting from the narrow angle approximation, not narrowed down by boundary conditions) and compare it with the family of experimental JONSWAP spectra.

As we will see, without the boundary conditions, the theoretical spectra form a family with 4 parameters, one parameter less than the 5 parameters in the JONSWAP spectral family. We will consider the three cases described in[4]: "Mean", "Sharp", and "Very sharp" JONSWAP spectra. In each of these situations, we will find parameters so that the theoretical spectrum agrees with the corresponding JONSWAP spectrum. However, I was unable to choose the four parameter to reproduce the Pierson-Moskowitz spectrum[27] with similar accuracy. Perhaps, this is related to the fact that the JONSWAP team[4] "have not been able to observe a systematic transition of the spectra into a fully developed spectrum of the form proposed by Pierson and Moskowitz". It seems that the Pierson-Moskowitz spectrum is insufficiently peaked.

Like in[9,18–20], we disregard certain logarithmic divergence. This does not seem as a serious neglect, as several other effects have been already disregarded; e.g. Hasselmann's kinetic equation disregards wave phase correlations; besides, additional terms in Hasselmann's equation for dissipation and forcing (wave excitation) are neglected. The logarithmic divergence is absent in Hasselmann's equation and only appears due to our approximation; but— like in[9,18–20] — this approximation greatly simplifies calculations and allows to see physical mechanism. Because of this disregard, certain quantitative predictions remain beyond the scope of the present paper.

## II. HASSELMANN'S EQUATION IN NARROW ANGLE APPROXIMATION

The turbulence of ocean surface waves is described by Hasselmann's kinetic equation[5,6]

$$\frac{\partial n_1}{\partial t} = \int R_{1234} \left[ n_3 n_4 (n_1 + n_2) - n_1 n_2 (n_3 + n_4) \right] \times$$
$$\delta(\mathbf{k}_1 + \mathbf{k}_2 - \mathbf{k}_3 - \mathbf{k}_4) \, \delta(\omega_1 + \omega_2 - \omega_3 - \omega_4) \, d_{234} \qquad (1)$$

for the wave action spectrum $n(\mathbf{k}, t)$. Here $\mathbf{k} = (p, q)$ is the wave vector, $\omega(\mathbf{k}) = \sqrt{g \, |\mathbf{k}|}$ is the dispersion ($g$ is the gravity acceleration);

$$\mathbf{k}_j = (p_j, q_j), \quad \omega_j = \omega(\mathbf{k}_j), \quad n_j = n(\mathbf{k}_j, t) \quad (j = 1, 2, \ldots);$$

integration is over 3 wave vectors $d_{234} = d\mathbf{k}_2 \, d\mathbf{k}_3 \, d\mathbf{k}_4$. The kernel $R_{1234} = R(\mathbf{k}_1, \mathbf{k}_2, \mathbf{k}_3, \mathbf{k}_4)$ is a complicated function; its exact expression was calculated and presented in different ways[5,28–31]; all expressions were shown to be consistent[23,32].

The present paper uses only the symmetries of the kernel:
*(1) symmetries with respect to transpositions of indices*

$$R_{1234} = R_{2134} = R_{1243} = R_{3412}; \quad (2a)$$

*(2) the mirror symmetry*

$$R(p_1, q_1, p_2, q_2, p_3, q_3, p_4, q_4) =$$
$$R(p_1, -q_1, p_2, -q_2, p_3, -q_3, p_4, -q_4); \quad (2b)$$

*(3) the scale homogeneity with degree 6*

$$R(\kappa \mathbf{k}_1, \kappa \mathbf{k}_2, \kappa \mathbf{k}_3, \kappa \mathbf{k}_4) = \kappa^6 R(\mathbf{k}_1, \mathbf{k}_2, \mathbf{k}_3, \mathbf{k}_4) \quad (2c)$$

for any positive number $\kappa$.

First, let us write Hasselmann's equation in a more symmetric form using an arbitrary test function $\chi(\mathbf{k})$: Multiply (1) by $\chi(\mathbf{k}_1)$, integrate over $d\mathbf{k}_1$, and interchange the integration variables, using the symmetries (2a)

$$\frac{\partial}{\partial t} \int n_1 \chi_1 \, d\mathbf{k}_1 = \frac{1}{4} \int R_{1234} \times$$
$$[n_3 n_4 (n_1 + n_2) - n_1 n_2 (n_3 + n_4)] (\chi_1 + \chi_2 - \chi_3 - \chi_4)$$
$$\times \delta(\mathbf{k}_1 + \mathbf{k}_2 - \mathbf{k}_3 - \mathbf{k}_4) \, \delta(\omega_1 + \omega_2 - \omega_3 - \omega_4) \, d_{1234} \quad (3)$$

($\chi_j = \chi(\mathbf{k}_j)$, $j = 1, 2, \ldots$). Notice factor $1/4$; it is due to the summation of four equal integrals resulting from the interchanges of the integration variables.

Let $p$ be the wave number along the main propagation direction ($p > 0$) and $q$ — the transverse wave number. We are interested in the quasi 1-D situation, when $|q| \ll p$. In exactly 1-D situation, when $q_1 = q_2 = q_3 = q_4 = 0$, the delta functions in equation (3) imply three possibilities: *(i)* $p_3 = p_1$ & $p_4 = p_2$, *(ii)* $p_4 = p_1$ & $p_3 = p_2$, or *(iii)* when one of the wave numbers has the opposite sign to the other wave numbers. However, in the third possibility, the kernel $R_{1234}$ turns into zero[33]. So, there is no energy transfer in the 1-D case; any stationary 1-D spectrum $n(p, q = 0)$ is always an exact solution of Hasselmann's equation. In the quasi 1-D situation ($q$ s are non-zero, but small), the energy exchange takes place:

$$p_3 \text{ is close to } p_1 \quad \& \quad p_4 \text{ is close to } p_2$$
$$\text{or} \quad p_4 \text{ is close to } p_1 \quad \& \quad p_3 \text{ is close to } p_2.$$

We can simplify calculations if from the beginning take into account the fact (noted in[4,7]) that the main contribution in the r.h.s. integral of Hasselmann's equation occurs when all four wave numbers $p_j$ are close to each other. Our calculations below will also support this fact. Introducing

$$u = p_3 - p_1 = p_2 - p_3, \quad v = p_4 - p_1 = p_2 - p_3,$$
$$\text{and} \quad p_0 = \frac{p_1 + p_2}{2} = \frac{p_3 + p_4}{2},$$

we have

$$p_1 = p_0 - \frac{u}{2} - \frac{v}{2}, \quad p_2 = p_0 + \frac{u}{2} + \frac{v}{2},$$
$$p_3 = p_0 + \frac{u}{2} - \frac{v}{2}, \quad p_4 = p_0 - \frac{h}{2} + \frac{v}{2}. \quad (4)$$

One can introduce (4) directly, considering any three of equations (4) as a change of variables (say, change from $p_1, p_2, p_3$ to $p_0, u, v$; its Jacobian is 1); then the fourth equation is due to the delta function $\delta(p_1 + p_2 - p_3 - p_4)$. We use (4) to expand quantities in Hasselmann's equation, assuming $|u|, |v| \ll p_0$. In these approximations, we take only the first non-trivial terms.

$$\frac{\omega(\mathbf{k})}{\sqrt{g}} \approx \sqrt{p} + \frac{q^2}{4p^{3/2}} \approx \sqrt{p_0} + \frac{p - p_0}{2\sqrt{p_0}} - \frac{(p - p_0)^2}{8p_0^{3/2}} + \frac{q^2}{4p_0^{3/2}}$$

(here the first approximation is due to $|q| \ll p$, and the second approximation uses $|p - p_0| \ll p_0$). Now substitute $p$ subsequently for $p_1, p_2, p_3, p_4$ and use (4)

$$\frac{\omega_1 + \omega_2 - \omega_3 - \omega_4}{\sqrt{g}} \approx \frac{-uv + q_1^2 + q_2^2 - q_3^2 - q_4^2}{4p_0^{3/2}}. \quad (5a)$$

We approximate the kernel $R_{1234}$ by its value when all its four wave vectors are equal

$$R_{1234} \approx R_0 \quad [\mathbf{k}_1 = \mathbf{k}_2 = \mathbf{k}_3 = \mathbf{k}_4 = (p_0, 0)], \quad (5b)$$

It is important that $R_0 \neq 0$[7]; otherwise, the approximation (5b) would be meaningless. According to the scaling symmetry (2c), $R_0 \propto p_0^6$. Actually, $R_0 = p_0^6/(4\pi^3)$, e.g.[31].

We do not expand $n$ with respect to $q \approx 0$ since the spectrum changes sharply in $q$, and so, the Taylor expansion in $q$ would provide a poor approximation. However, we approximate the spectra by their values when $p_j \approx p_0$

$$n_j \approx n(p_0, q_j, t) \quad (j = 1, 2, 3, 4). \quad (5c)$$

Like in[9,18–20], we restrict ourselves to symmetric spectra

$$n(p, q, t) = n(p, -q, t). \quad (6)$$

Since the medium possesses the mirror symmetry (2b), then if the condition (6) holds initially, it will hold forever. The condition (6) implies that all odd moments of the spectrum are zero

$$\int q \, n(p, q, t) \, dq = \int q^3 \, n(p, q, t) \, dq = \ldots = 0.$$

We will derive coupled equations for the 1-D spectrum and the second moment:

$$N(p, t) = \int n(p, q, t) \, dq, \quad Z(p, t) = \int q^2 \, n(p, q, t) \, dq. \quad (7)$$

To derive these equations, we will use two types of test functions $\chi(p, q)$.

**A. Equation for the 1-D spectrum**

First, take the test function independent of $q$

$$\chi(p, q) = \phi(p) \quad \Rightarrow \quad \chi_1 + \chi_2 - \chi_3 - \chi_4 \approx \phi_0'' u v, \quad (8)$$

where $\phi(p)$ is an arbitrary 1-D test function, $\phi_0''$ is the value of the second derivative $\phi''(p)$ at $p = p_0$. Then with the approximations (5abc) and (8), the equation (3) takes the form

$$\frac{\partial}{\partial t} \int N_1 \phi_1 \, dp_1 = \int R_0 [n_3 n_4 (n_1 + n_2) - n_1 n_2 (n_3 + n_4)]$$
$$\times \phi_0'' \, u \, h \, \frac{p_0^{3/2}}{g^{1/2}} \, \delta(-uv + q_1^2 + q_2^2 - q_3^2 - q_4^2) \times$$
$$\delta(q_1 + q_2 - q_3 - q_4) \, dp_0 \, du \, dv \, dq_1 \, dq_2 \, dq_3 \, dq_4. \quad (9)$$

The integration over $du$ at the expense of the delta function gives

$$uv = q_1^2 + q_2^2 - q_3^2 - q_4^2$$

and the Jacobian factor $|v|^{-1}$. Now, because of approximation (5c), we can explicitly integrate over $dq_1\, dq_2\, dq_3\, dq_4$. For example, for the first term in the brackets of (9)

$$\int n_3\, n_4\, n_1\, (q_1^2 + q_2^2 - q_3^2 - q_4^2)\, \delta(q_1 + q_2 - q_3 - q_4) \times$$
$$dq_1\, dq_2\, dq_3\, dq_4$$
$$\approx \int n(p_0, q_3)\, n(p_0, q_4)\, n(p_0, q_1) \times$$
$$[q_1^2 + (q_3 + q_4 - q_1)^2 - q_3^2 - q_4^2]\, dq_1\, dq_3\, dq_4 = 2N_0^2 Z_0. \quad (10)$$

Here we took into account the spectral symmetry (6), so that the only non-zero contribution from the bracket in (10) is due to the term $2q_1^2$. Similar, we integrate other terms in the bracket of (9) and find

$$\frac{\partial}{\partial t} \int N_1 \phi_1\, dp_1 = K \int p_0^{15/2} N_0^2\, Z_0\, \phi_0''\, dp_0 \quad (11)$$

with the "constant" that includes logarithmic divergence

$$K = \frac{4}{\pi^3}\, g^{-1/2} \int \frac{dv}{|v|}. \quad (12)$$

One can consider $K$ as a fitting constant.

Now, integrating by parts in the r.h.s. of (11) and using the arbitrariness of the function $\phi(p)$, we obtain the evolution equation for the 1-D spectrum $N(p,t)$

$$\frac{\partial N}{\partial t} = K \frac{\partial^2}{\partial p^2} \left( p^h\, N^2\, Z \right) \qquad \left( h = \frac{15}{2} \right). \quad (13)$$

### B. Equation for the second moment

As a second type of test function, take it proportional to $q^2$

$$\chi(p,q) = \psi(p)\, q^2 \quad \Rightarrow$$
$$\chi_1 + \chi_2 - \chi_3 - \chi_4 \approx \psi_0\, (q_1^2 + q_2^2 - q_3^2 - q_4^2) \quad (14)$$

where $\psi(p)$ is an arbitrary 1-D test function, $\psi_0$ is its value at $p = p_0$. Now we reason similar to the previous subsection. With the approximations (5abc) and (14), the equation (3) takes the form

$$\frac{\partial}{\partial t} \int Z_1 \psi_1\, dp_1 = \int R_0\, [n_3 n_4 (n_1 + n_2) - n_1 n_2 (n_3 + n_4)]$$
$$\times \psi_0\, (q_1^2 + q_2^2 - q_3^2 - q_4^2)\, \frac{p_0^{3/2}}{g^{1/2}}\, \delta(-uv + q_1^2 + q_2^2 - q_3^2 - q_4^2)$$
$$\times \delta(q_1 + q_2 - q_3 - q_4)\, dp_0\, du\, dv\, dq_1\, dq_2\, dq_3\, dq_4. \quad (15)$$

Here the r.h.s. integrand depends on $u$ only through the delta function, and the integration over $du$ just gives the Jacobian factor $|v|^{-1}$. Using approximation (5c), we integrate over $dq_1\, dq_2\, dq_3\, dq_4$. Similar to calculation (10), we find

$$\frac{\partial}{\partial t} \int Z_1 \psi_1\, dp_1 = K \int p_0^{15/2} N_0^2\, Z_0\, \psi_0\, dp_0 \quad (16)$$

with the same constant $K$, given in (12). Since the function $\psi(p)$ is arbitrary, we obtain the evolution equation for the second moment $Z(p,t)$

$$\frac{\partial Z}{\partial t} = K p^h\, N^2\, Z \qquad \left( h = \frac{15}{2} \right). \quad (17)$$

Equation (17) implies that the second moment is always increasing, a conclusion found in[9]. Although we used a different method, we arrived at the same conclusion $\partial Z/\partial t > 0$, which means that the turbulence cannot evolve to a stationary spectrum with narrow angular distribution: The spectrum will always be broadening in angle.

The equations (13) and (17) form a coupled system for the functions (7). These equations imply

$$\frac{\partial N}{\partial t} = \frac{\partial^2}{\partial p^2} \frac{\partial Z}{\partial t} \quad \Longleftrightarrow \quad N = \frac{\partial^2 Z}{\partial p^2} + X(p) \quad (18)$$

with some function $X(p)$ independent of time. Substituting (18) into (17), we arrive at a single equation

$$\frac{\partial Z}{\partial t} = K p^h \left( \frac{\partial^2 Z}{\partial p^2} + X \right)^2 Z, \quad (19)$$

which contains an arbitrary time-independent function $X(p)$. The solution $Z(p,t)$ determines the 1-D spectrum, by (18). By their definition (7), $N > 0$ and $Z > 0$ $(0 < p, t < \infty)$.

The present paper considers the simplest case when $X \equiv 0$. We can also eliminate $K$ by re-scaling time: $t \to t/K$. Thus,

$$N = \frac{\partial^2 Z}{\partial p^2}, \qquad \frac{\partial Z}{\partial t} = p^h \left( \frac{\partial^2 Z}{\partial p^2} \right)^2 Z. \quad (20)$$

## III. SOLUTIONS

### A. First approximation

Since $h$ is relatively large, we can solve equations (20) by iterations, re-writing them in the form

$$N = \frac{\partial^2 Z}{\partial p^2} = p^{-h/2} \sqrt{\frac{1}{Z} \frac{\partial Z}{\partial t}} \quad (21)$$

For large $p$, the spectrum should be small, and so, as zeroth order approximation, we take $N = 0$. Then the second moment is a linear function $Z = C(t)\, p + \tilde{C}(t)$, and

$$N = p^{-h/2} \sqrt{\frac{\dot{C} p + \dot{\tilde{C}}}{C p + \tilde{C}}}, \quad (22a)$$

This can be written in the form with 3 parameters

$$N = B p^{-h/2} \sqrt{\frac{p + b}{p + c}} \quad (22b)$$

($B, b, c$ are parameters depending on time). Let us compare this spectrum with the Pierson-Moskowitz spectrum

$$E_\omega^{\text{P-M}} = A \left( \frac{\omega}{\omega_m} \right)^{-5} \exp\left[ -\frac{5}{4} \left( \frac{\omega}{\omega_m} \right)^{-4} \right]. \quad (23)$$

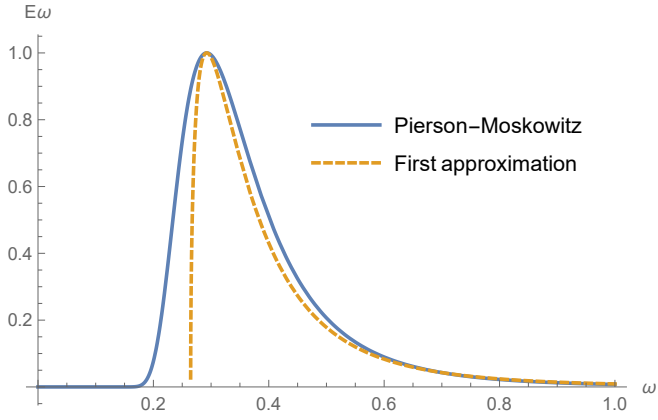

FIG. 1: The first approximation spectrum (25b) vs. the Pierson-Moskowitz spectrum (23); $b = -0.07$, $\omega_m = 0.3$.

The energy spectrum $E_\omega$ with respect to the frequency is related to the 1-D wave action spectrum $N$ by the relation

$$E_\omega \, d\omega = \omega N \, dp \qquad (p = \omega^2). \tag{24}$$

So, the wave action spectrum (22b) translates into the energy spectrum

$$E_\omega = 2\omega^2 N = 2B\,\omega^{2-h} \sqrt{\frac{\omega^2 + b}{\omega^2 + c}}. \tag{25a}$$

This spectrum has a peak when $b < 0$ and $c > 0$. The shape of the spectrum is almost insensitive to the value of $c > 0$. A slightly better fit to the Pierson-Moskowitz spectrum is achieved for large $c$, so that the spectrum (25a) essentially has two parameters

$$E_\omega = 2B\,\omega^{2-h}\sqrt{\omega^2 + b}. \tag{25b}$$

(here $B$ differs from $B$ in (22b) and (25a)).

We can choose parameters so that this spectrum and the Pierson-Moskowitz spectrum have the same maximum value reached at the same frequency $\omega_m$. The Figure 1 shows the two spectra. The JONSWAP spectra are known to be significantly more peaked than the Pierson-Moskowitz spectrum; and so, there is no way the first approximation spectrum (22b) could approximate the JONSWAP spectra. A better approximation is reached by considering self-similar solutions of equations (20).

### B. Self-similar spectra

The present paper advocates that the spectral anisotropy grows only at the down-shifting peak (when the wave number $p$ is around its value $p_m$ at the peak). Behind the peak (when $p \gg p_m$), the spectrum is broadening in angle. So, it makes sense to look for self-similar solutions

$$N(p,t) = t^\nu F(pt^\lambda), \qquad Z(p,t) = t^\zeta G(pt^\lambda) \tag{26}$$

with some unknown exponents $\lambda$, $\nu$, $\zeta$ and "shape" functions $F(x)$, $G(x)$  $(0 < x < \infty)$.

Substituting (26) into equations (20), we find relations for the exponents

$$\nu = \zeta + 2\lambda, \qquad h\lambda - 2\nu = 1 \qquad (h = 15/2) \tag{27}$$

and the second-order o.d.e. for the shape function

$$\zeta G + \lambda x G' = x^h (G'')^2 G, \tag{28a}$$

which can be solved for the second derivative

$$G'' = x^{-h/2}\sqrt{\zeta + \lambda x G'/G}. \tag{28b}$$

Solutions $G(x)$ determine the 1-D spectra: $F(x) = G''(x)$.

We will see that the equation (28) indeed leads to the spectral peak, typical for the JONSWAP spectrum; it abruptly decreases on the forward face of the peak and gradually decreases behind the peak.

As we can see from (26), the spectral peak would down-shift if $\lambda > 0$. The spectral anisotropy at the peak would grow if $\nu > \zeta$. The latter two conditions are equivalent, as we can see from the first equation in (27). So, the down-shift is equivalent to the anisotropy growth.

Now let us compare the theoretical spectrum, resulting from the o.d.e. (28), with the experimental JONSWAP[4] spectrum

$$E_\omega^{\text{JONSWAP}} = E_\omega^{\text{P-M}} \gamma^{\exp\frac{-(\omega-\omega_m)^2}{2\sigma^2 \omega_m^2}}, \tag{29}$$

$$\sigma = \begin{cases} \sigma_a, & \omega \leq \omega_m, \\ \sigma_b, & \omega > \omega_m, \end{cases}$$

$\gamma, \sigma_a, \sigma_b$ are three more fitting parameters in addition to parameters $A, \omega_m$ of the Pierson-Moskowitz spectrum (23).

According to the relation (24), the theoretical energy spectrum $E_\omega$ with respect to the frequency is

$$E_\omega = 2x\,G''(x)\,t^{\nu-\lambda}, \tag{30}$$

where $x = \omega^2 t^\lambda$.

The equation (28) possesses a *continuous symmetry*: If $G(x)$ is a solution of this equation, then the function

$$\tilde{G}(x) = a^{2-h/2}\,G(x/a) \tag{31}$$

is also a solution of (28), for any positive constant $a$. This symmetry allows to control the magnitude of the theoretical spectrum (30) — even though the equation (28) is nonlinear — and in particular, to have $E_\omega$ of the same magnitude as the experimental spectrum (29), with an arbitrary magnitude, controlled by parameter $A$.

Actually, we can choose $a$ and $t$, so that the theoretical spectrum $E_\omega$ and the experimental spectrum (29) reach the same maximum value at the same frequency

$$E_\omega(\omega_m) = E_\omega^{\text{JONSWAP}}(\omega_m) = 1. \tag{32}$$

We have chosen $A$ in (29) to have the JONSWAP spectrum with unit magnitude. Let us also choose $\omega_m = 1/3$.

The solutions $G(x)$ of (28) are uniquely determined by two initial conditions

$$G(x_m) = G_m, \qquad G'(x_m) = G'_m;$$

here for the initial point we take the point $x_m$ where the function $x G''(x)$ has maximum (tip of the peak, see (30)). So, the spectrum (30) is determined by four parameters: $\lambda$, $G_m$, $G'_m$ and $t$ (recall that $t$ is time re-scaled by $K$). Two parameters determine the location of the peak and its magnitude, and two — the shape of the peak. The parameter choice is detailed in Appendix.

Figure 2 shows the comparison of the theoretical spectra (30) with the experimental spectra (29) in the three cases specified in Ref.[4] as "Mean", "Sharp", and "Very sharp" JONSWAP spectra. The figure shows that in each of the three cases, we can choose the four parameters to approximate the JONSWAP spectrum by the corresponding theoretical spectrum. The biggest difference is on the forward face of the spectral peak in the part where the spectrum is relatively small, see Fig. 2. Such difference (small values over the small interval of sharp forward face) contributes little to the integral characteristics of the turbulence.

Figure 3 shows the typical graphs of the functions $F(x)$ and $G(x)$. We see that the graphs abruptly stop at some $x < x_m$. This always happens when $\zeta < 0$, and the expression under the square root in the equation (28b) can become negative with decreasing $x$. In these situations, the second moment $Z(p,t)$ is decreasing as we move with the peak. Thus, the spectrum does not extend to the boundary $\omega \to 0$.

On the opposite end, the spectrum automatically vanishes as $\omega \to \infty$. Herewith, the second moment $Z$ linearly grows, see Fig. 3; this means that the spectrum becomes isotropic, and the narrow angle approximation fails before the boundary is reached. So, the imposition of boundary conditions does not lead to determination of some parameters in the spectral family.

## IV. CONCLUSION

We have considered the narrow angle approximation for Hasselmann's kinetic equation (Sect. II), resulting in a differential equations (20) for the second moment $Z(p,t)$, defined in (7). The approximation is a differential equation, but not for the spectrum. The present paper re-confirms the finding in Ref.[9] that for any fixed $p$, the second moment $Z(p,t)$ grows, and the spectrum is broadening in angle. Only as we move with the downshifting peak, we see the growth of spectral anisotropy. On the backward face of the downshifting peak, the spectrum gradually becomes isotropic.

This behavior is described by self-similar solutions (26) with exponents $\lambda, \nu, \zeta$, satisfying the relations (27), and the shape function $G(x)$, determined by the 2-nd order o.d.e. (28). Its solution determines not only $Z(p,t)$, but also the 1-D spectrum $N(p,t)$: $F(x) = G''(x)$. So, the family of theoretical spectra depends on 4 parameters (2 initial conditions, along with $\lambda$ and $t$). The exact parameter choice is described in appendix A.

As we discussed at the end of Section III B, the imposition of boundary conditions would not reduce the number of parameters. Moreover, the narrow angle approximation (Section II) fails before the boundaries are reached. We compared

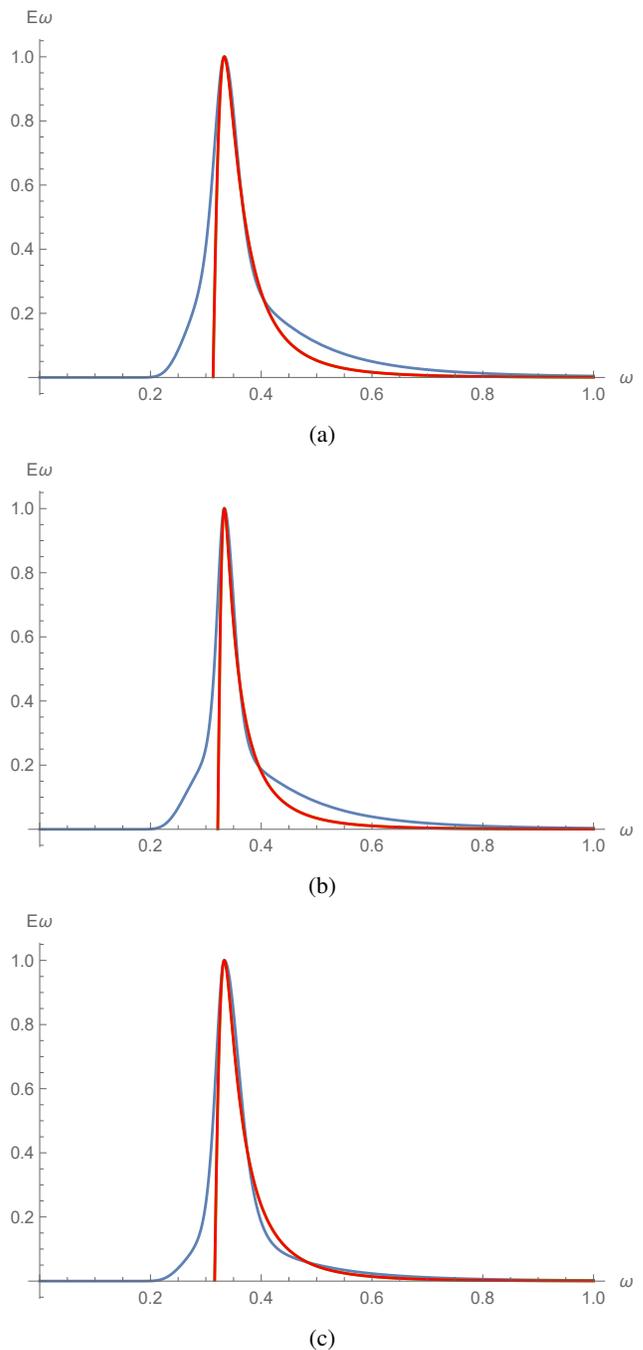

FIG. 2: The experimental JONSWAP spectrum[4] (blue line) and the theoretical spectrum $E_\omega$ (red line).
(a) "Mean" JONSWAP $\gamma = 3.3$, $\sigma_a = 0.07$, $\sigma_b = 0.09$, $\lambda = \frac{2}{9}$, $\nu = \frac{1}{3}$, $\zeta = -\frac{1}{9}$, $t = 1.0 \times 10^6$, $G_m = 1.5$, $G'_m = 96$.
(b) "Sharp" JONSWAP $\gamma = 4.2$, $\sigma_a = 0.05$, $\sigma_b = 0.07$, $\lambda = \frac{2}{9}$, $\nu = \frac{1}{3}$, $\zeta = -\frac{1}{9}$, $t = 2.4 \times 10^6$, $G_m = 0.57$, $G'_m = 82$.
(c) "Very sharp" JONSWAP $\gamma = 7.0$, $\sigma_a = 0.07$, $\sigma_b = 0.12$, $\lambda = \frac{2}{9}$, $\nu = \frac{1}{3}$, $\zeta = -\frac{1}{9}$, $t = 1.3 \times 10^6$, $G_m = 1.2$, $G'_m = 93$.
6

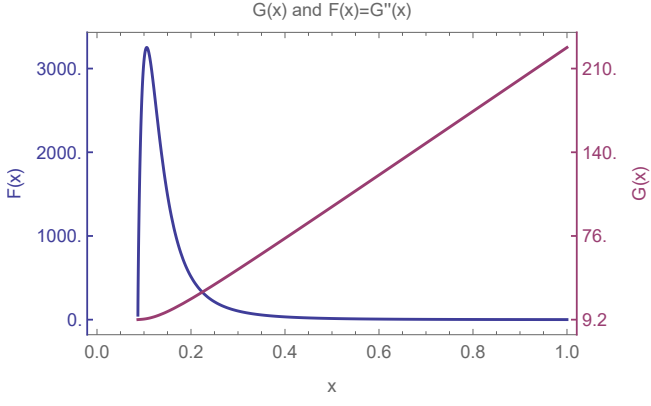

FIG. 3: The shape functions $F(x)$ and $G(x)$ defining the self-similar solutions (26). Like in the previous figure, $\lambda = \frac{2}{9}$, $\nu = \frac{1}{3}$, $\zeta = -\frac{1}{9}$.

the entire 4-parameter family of theoretical spectra with established experimental JONSWAP spectra.

The agreement appears satisfactory. The biggest difference occurs on the forward face in the lower part of the peak and seems insignificant for the integral turbulence properties. At the same time, I was unable to match the theoretical spectrum to the Pierson-Moskowitz spectrum[27] with similar accuracy. This might indicate that the 5-parameter JONSWAP spectral family contains physically unrealizable spectra.

**Appendix A: Parameters of the theoretical spectra**

Let us define the parameter

$$z = \frac{\zeta}{\lambda} \tag{A1}$$

and consider solutions $H(y)$ of the initial value problem

$$H'' = y^{-h/2}\sqrt{z + yH'/H}, \tag{A2a}$$
$$H(y_m) = H_m,\ H'(y_m) = H'_m. \tag{A2b}$$

Solutions $G(x)$ of equation (28) are re-scaled solutions $H(y)$ of (A2a)

$$G(x) = a^{2-h/2}\sqrt{\lambda}\, H\left(\frac{x}{a}\right). \tag{A3}$$

Here the square root is well defined because $\lambda > 0$, the condition of the down-shift. One can think of the variable $y$ as dimensionless version of $x$: $y = x/a$ where variable $x$ and positive constant $a$ have the same units, cf. (31). Similar to $x_m$ chosen in Section III B, for the initial point of initial conditions (A2b), we take the point $y_m$ where the function $y\,H''(y)$ has maximum.

Substitution of (A3) into (30) gives

$$E_\omega = 2y\,H''(y)\left(\frac{a}{t^\lambda}\right)^{1-h/2}\sqrt{\frac{\lambda}{t}}, \tag{A4a}$$

where $y = \omega^2 t^\lambda/a$. Here we took into account the expression of $\nu$ via $\lambda$, which follows from (27).

Let the function $f(y) = y\,H''(y)$ reach some max value $M$ at $y = y_m$. Then, by (A4a), $E_\omega$ reaches maximum value

$$2M\,(a/t^\lambda)^{1-h/2}\sqrt{\frac{\lambda}{t}},$$

at point $\omega = \omega_m$ such that $y_m = \omega_m^2 t^\lambda/a$.

Like the equation (28), the equation (A2a) posses the continuous symmetry similar to (31). Therefore, we can select the particular solution $H$ so that the function $f(y) = y\,H''(y)$ has maximum at any given point. Let us select $H(y)$ with maximum at point $y = y_m = \omega_m^2$. Then $a = t^\lambda$ and the condition (32) holds when $t = 4\lambda\,M^2$; herewith the formula (A4a) simplifies to

$$E_\omega = y\,H''(y)/M, \qquad y = \omega^2. \tag{A4b}$$

The maximum condition $f'(y_m) = 0$ connects $y_m$ with initial values (A2b). One can calculate $f'(y)$ without finding $H(y)$: After differentiating, substitute $H''$ from the differential equation (A2a). So,

$$H_m = \frac{y_m^{2-h/2}\sqrt{R_m + z}}{R_m^2 + (h-3)R_m + (h-2)z}, \tag{A5a}$$
$$H'_m = H_m\,R_m/y_m, \tag{A5b}$$

Here we have introduced dimensionless ratio

$$R(y) = \frac{y\,H'(y)}{H(y)}, \qquad R_m = R(y_m). \tag{A6}$$

The two parameters $z$ and $R_m$ control the shape of the theoretical spectrum $E_\omega$, while the remaining two parameters $a$ and $t$ control the location of the peak and its magnitude.

The relations (27) with (A1) determine the self-similarity exponents in terms of $z$

$$\lambda = \frac{2}{7-4z}, \qquad \zeta = z\lambda, \qquad \nu = (z+2)\lambda. \tag{A7}$$

The peak would down-shift if

$$\lambda > 0 \quad \Leftrightarrow \quad z < 7/4. \tag{A8a}$$

Under the requirement (A8a), the peak would grow if

$$\nu > 0 \quad \Leftrightarrow \quad z > -2 \quad \Leftrightarrow \quad \lambda > 2/15, \tag{A8b}$$

while the second moment $Z(p,t)$ would decrease as we move with the peak if

$$\zeta < 0 \quad \Leftrightarrow \quad z < 0 \quad \Leftrightarrow \quad \lambda < 2/7. \tag{A8c}$$

Although the parameter $z$ determines the self-similarity exponents, and in particular, the downshift of the spectral peak, I found no significant dependence on $z$ of the shape of the peak. The figures 2 and 3 all use $z = -1/2$.